\documentclass[9pt]{revtex4}
\usepackage{hyperref}
\usepackage{datetime}
\usepackage{amsmath,amsfonts,amsthm,amssymb}
\usepackage{mdframed}
\usepackage{cases}
\usepackage{color}
\usepackage{graphicx,subcaption,float,wrapfig}
\usepackage{overpic}
\usepackage[usenames, dvipsnames]{xcolor}
\usepackage{algorithm}
\usepackage{multirow}
\usepackage{dcolumn}
\usepackage{bm}
\usepackage[noend]{algorithmic}
\renewcommand{\algorithmicrequire}{\textbf{Input:}}
\renewcommand{\algorithmicensure}{\textbf{Output:}}
\usepackage{amsmath,amssymb,amsfonts,mathrsfs,caption,epstopdf,mathptmx,hyperref}
\usepackage[capitalise]{cleveref}
\hypersetup{colorlinks=true,linkcolor=Blue,citecolor=Blue}
\numberwithin{equation}{section}
\begin{document}

\title{Solution Landscapes of the Simplified Ericksen--Leslie Model and its Comparison with the Reduced Landau--de Gennes Model}

\author{
Yucen Han$^{1, 4}$, Jianyuan Yin$^{2}$, Yucheng Hu$^{3}$, Apala Majumdar$^{1}$, Lei Zhang$^{4,5}$}

\address{$^{1}$Department of Mathematics and Statistics, University of Strathclyde, G1 1XQ, UK.\\
$^{2}$School of Mathematical Sciences, Peking University, Beijing 100871, China.\\
$^{3}$Beijing Advanced Innovation Center for Imaging Theory and Technology, Academy for Multidisciplinary Studies, Capital Normal University, Beijing 100048, China.\\
$^{4}$Beijing International Center for Mathematical Research, Peking University, Beijing 100871, China.\\
$^{5}$Center for Quantitative Biology, Peking University, Beijing 100871, China.}

\begin{abstract}
We investigate the solution landscapes of a simplified Ericksen--Leslie (sEL) vector model for nematic liquid crystals, confined in a two-dimensional square domain with tangent boundary conditions.
An efficient numerical algorithm is developed to construct the solution landscapes by utilizing the symmetry properties of the model and the domain.
Since the sEL model and the reduced Landau--de Gennes (rLdG) models can be viewed as Ginzburg--Landau functionals, we systematically compute the solution landscapes of the sEL model, for different domain sizes, and compare with the solution landscapes of the corresponding rLdG models. 
There are many similarities, including the stable diagonal and rotated states, bifurcation behaviors, and sub-solution landscapes with low-index saddle solutions. Significant disparities also exist between the two models. The sEL vector model exhibits the stable solution $C\pm$ with interior defects, high-index ``fake defects" solutions, novel tessellating solutions, and certain types of distinctive dynamical pathways. The solution landscape approach provides a comprehensive and efficient way for model comparison and is applicable to a wide range of mathematical models in physics.
\end{abstract}
\maketitle
\section{Introduction}
Nematic liquid crystals (NLCs) are the simplest type of liquid crystals, such that the constituent asymmetric NLC molecules have no translational order but exhibit a degree of long-range orientational order \cite{de1993physics}. Consequently, NLCs have distinguished material directions, referred to as \emph{nematic directors}, and direction-dependent properties i.e. the material properties along the directors are different from the properties in other directions. NLCs are classical examples of partially ordered materials and NLC order is ubiquitous in nature e.g. one can see imprints of NLC order in human DNA, the cell cytoskeleton, bacterial suspensions, flocks of moving animals, polymers, colloids to name a few. Hence, the modeling and computational methods for NLCs can be applied beyond the specific field of NLCs, e.g. for allied fields such as active matter, elastomers, and biomimetic materials. One of the most important features of NLCs is the topologically induced defects with beautiful and fascinating optical fingerprints \cite{ball2017liquid,machon2016global}. These defects can be interpreted as discontinuities in nematic directors or localised regions of loss of NLC orientational order. NLC defects can be classified according to their topological degrees, e.g. $\pm1$ and $\pm1/2$ point defects in two-dimensions (2D). NLC defects are energetically unfavorable due to the additional elastic energy around them and yet they are unavoidable due to the geometric frustration, topological reasons and external fields, and as we will see, they play a crucial role in the solution landscapes to determine the stability or degree of instability of NLC states in confinement.

There are different mathematical theories for NLCs with differing levels of details, such as the mean-field Maier--Saupe theory and Onsager theories \cite{maier1958einfache,onsager1949effects}, macroscopic continuum theories including the Oseen--Frank theory \cite{oseen1933theory,frank1958liquid}, the Ericksen--Leslie (EL) theory \cite{ericksen1991liquid}, and the celebrated Landau-de Gennes (LdG) $\mathbf{Q}$-tensor theory \cite{de1993physics}. The continuum theories do not explicitly include molecular-level information and are based on the assumption that NLC structural properties vary slowly on molecular length scales i.e. the molecular details are averaged out for macroscopic observables. The continuum/macroscopic theories are defined by a macroscopic order parameter that is a measure of the degree of orientational order or state of NLC anisotropy. There are multiple ways for choosing the macroscopic order parameters and the relationships between the different continuum theories have been extensively studied in \cite{han2015microscopic,ball2010nematic,ball2017liquid,wang2021modeling}.
In this manuscript, we focus on two continuum theories --- the EL and LdG theories, in the 2D square as a specific example.  As a vector model, the EL theory is restricted to uniaxial NLCs, with a single distinguished nematic director described by a unit-vector field, $\bm n$, that models the single locally preferred orientation direction of the NLC molecules at each point in space. 
The degree of orientational order about $\bm n$ is described by a scalar order parameter, $s$, and in particular, $s$ vanishes at NLC defects. The EL model is limited in the sense that it does not respect the head-to-tail symmetry of NLC states ($\bm n$ is equivalent to $-\bm n$ for rod-like molecules), which consequently leads to the defect lines between antiparallel directions and missing states containing interior half integer point defects, where the director rotates by $k\pi$-radians around the defect with $an$ odd number $k$ \cite{ball2017liquid}. Moreover, the EL model cannot describe biaxial NLC states with a primary and secondary NLC director. There are a number of theoretical works on the existence and regularity of energy minimizing configurations and defects in the EL model \cite{1991Nematic,2001Static}. The EL theory has also been used to study NLC flow dynamics \cite{leslie1979theory,lin1989nonlinear} and one can read about numerical schemes for the EL model in \cite{lin1996partial,badia2011overview}.

The LdG theory is one of the most powerful continuum theories for NLCs, that circumvents the limitations cited above \cite{de1993physics}. The macroscopic LdG order parameter is labelled as the $\mathbf{Q}$-tensor order parameter, which is a symmetric traceless $3\times 3$ matrix that respects the head-to-tail symmetry of NLC molecules. The eigenvectors of $\mathbf{Q}$ encode the nematic directors, and the corresponding eigenvalues measure the degree of orientational/directional order about the eigenvectors. In particular, there can be two scalar order parameters, associated with a primary and secondary director, in the LdG setting. 
The LdG theory is hugely successful in the context of predicting NLC phase transitions and rheological parameters, such as NLC viscosity and stress tensors, and has been extensively applied in various NLC systems in confinement \cite{han2019transition,hu2016disclination}.
Recent years have seen a boom in LdG studies of NLCs in confined 2D and three-dimensional (3D) settings, to predict experimentally observable static scenarios and dynamical pathways between distinct NLC states. For example, NLCs on square domains have received substantial attention in the LdG framework. In particular, for small nano-scale square domains, there is a unique well order reconstruction solution (WORS) with a pair of defect lines along the square diagonals \cite{kralj2014order}. The defect lines partition the square domain into quadrants such that the director is constant in each quadrant, and there are director jumps across the defect lines. The WORS exists for all square sizes but loses stability as the square edge length increases  \cite{canevari2017order}. For large square domains (on the micron scale or larger), there are two competing stable physically observable states: the largely uniaxial diagonal ($D$) states for which the nematic director is aligned along one of the square diagonals, and the rotated ($R$) states for which the uniaxial director rotates by $\pi$ radians between a pair of opposite edges \cite{kusumaatmaja2015free}.

Despite extensive studies for NLCs on 2D square domains in the LdG model, there are relatively few mathematical studies in the EL framework due to its limitations, e.g. orientability issues since the EL model is restricted to vector fields and inability to capture biaxiality \cite{henao_majumdar_pisante}. 
Thus, it is important to understand what the EL model can capture in terms of the admissible states, the defect configurations and the analogies and differences between the EL and LdG theories in confinement.
To address these questions, we apply a solution landscape approach to investigate a simplified Ericksen--Leslie (sEL) model on a 2D square domain, and compare it with a reduced Landau--de Gennes (rLdG) model. In a 2D setting such as ours, both models reduce to the pioneering Ginzburg--Landau (GL) model, which is a variational theory for superconductors \cite{bethuel1994ginzburg}. The GL model is amongst the most widely studied in the calculus of variation communities, and we expect two simplified models to share stable defect configurations, at least in some geometrical parameter regimes. 
The sEL and rLdG models are variational models i.e. the physically observable states are modelled by stable critical points (local/global minimizers) of the associated free energy, which is just a rescaled version of the GL energy. The critical points are solutions of the corresponding Euler-Lagrange equations, which are just the GL equations --- a system of two nonlinear, coupled elliptic partial differential equations for two independent components of the respective order parameters.

Mathematically, there are two essential differences in the sEL approach and the rLdG approach in this paper. The first difference concerns the order parameters and the concomitant orientability issues.
The sEL order parameter is a 2D vector $\mathbf{p}$, and the rLdG order parameter is a symmetric, traceless $2\times 2$ matrix with two independent components.
The second crucial difference is that we are solving two different boundary-value problems for the GL system, in the sEL and rLdG settings respectively.  We impose tangent boundary conditions in both cases, which require the nematic director to be tangent to the square edges necessarily creating defects at the square vertices. However, we use Dirichlet conditions for $\mathbf{Q}$ in the rLdG setting, and we only impose Dirichlet conditions for $\mathbf{n}$ in the sEL setting, keeping the order parameter $s$ free. Hence, the boundary conditions are much weaker (by choice) in the sEL setting. The purpose of this weak boundary condition is to recover the experimentally and theoretically reported $D$ and $R$ solutions. Notably, the isotropic state with $\mathbf{p} = \mathbf{0}$ is a solution of the sEL problem, whereas the isotropic state is not a critical point of the rLdG problem in our setting. This inherently means that the sEL problem can accommodate more exotic and numerous defect structures, at least for our choice of the boundary conditions.

There are multiple critical points in both models, including stable/metastable states, i.e., global/local minima, and unstable saddle points of the sEL and rLdG boundary-value problems. 
According to Morse theory \cite{milnor1963morse}, the saddle points can be classified by their Morse index. The (Morse) index of a critical point is the maximal dimension of a subspace on which the Hessian operator is negative definite i.e. the number of negative eigenvalues of the Hessian of the free energy which captures the number of unstable directions for an unstable saddle point.
Compared to numerical computations of stable states by gradient flow dynamics, saddle points are much harder for computational purposes and yet saddle points play a critical role in selection mechanisms and switching mechanisms between distinct stable critical points. For instance, the transition pathways mediated by an index-$1$ saddle point between $D$ and $R$ states have been studied in \cite{kusumaatmaja2015free}.
In order to efficiently compute both stable states and unstable saddle points, we construct the solution landscape as a pathway map from the high-index saddle points, to low-index saddle points and then index-$0$ energy minimizers \cite{yin2020construction}. The solution landscape is numerically constructed by combining the saddle dynamics (SD) with downward/upward search algorithms \cite{yin2019high,yin2020searching}, to connect high-index saddles to low-index ones and equally connect low-index saddles to high-index ones. The solution landscape approach has been successfully applied to the rLdG models on 2D squares \cite{yin2020construction} and hexagons \cite{han2020SL}, and to the 3D molecular Onsager model with different interaction kernel potentials \cite{yin2021solution}. 

In this manuscript, we develop a numerical algorithm that exploits the symmetry properties of the sEL and rLdG models to efficiently construct the corresponding solution landscapes. This is the key to reducing computational costs and improving efficiency of numerical algorithms, and is novel in the NLC context, to the best of our knowledge. We compare the solution landscapes for the sEL and rLdG models for different domain sizes, to capture the effects of geometrical size on the solution landscapes. As expected, the orientable nature of the sEL order parameter enforces new additional critical points, not found in the rLdG setting. Further, the weaker implementation of the tangent boundary conditions in the sEL setting, and the absence of constraints on $s$, necessarily implies that we find a plethora of defective critical points, with high indices, in the sEL setting. This includes saddle points with multiple interior defects, defects near vertices, edges and in the interior and novel tesselating saddle points. Notably, we find a new stable $C\pm$ state with an interior $\pm 1$-vortex at the square centre. The $C\pm$ state is found as an unstable saddle in the rLdG setting, since the non-orientable nature of the rLdG order parameter allows for the splitting of the interior vortex into pairs of $\pm \frac{1}{2}$-interior defects. This paper illustrates the interplay between the choice of the order parameter, the boundary conditions and the geometrical size on solution landscapes to some extent, and we hope that our work will trigger further substantive studies on these lines. 

The paper is organized as follows.
In Section \ref{sec:models}, we briefly introduce the sEL and rLdG models and discuss the relationships between them.
In Section \ref{sec:algorithm_construction}, we propose an efficient numerical algorithm to construct the solution landscapes such that the algorithms incorporate symmetry properties.
In Section \ref{sec:results}, we systematically construct the solution landscapes of the sEL model with an increasing  domain-size parameter, and compare with the rLdG predictions. 
Finally, we present the conclusions and discussions in Section \ref{sec:discussion}.

\section{Models}\label{sec:models}
\subsection{Ericksen--Leslie model}
The derivation of the EL model traces back to the Oseen--Frank (OF) model \cite{oseen1933theory}, a classical vector model in NLC theory.
In the OF model, the NLC states is assumed to be purely uniaxial with constant orientational order, and the OF order parameter is a unit-vector field, $\mathbf{n}=(n_1, n_2, n_3)^\top \in \mathbb{S}^{2}$, that models the locally-preferred averaged direction of alignment of the NLC molecules at each point in space. The OF theory is also a variational theory, so physically observable states are modelled by stable critical points, i.e. local/global minimizers of an appropriately defined OF free energy. The OF energy density is quadratic in $\nabla \mathbf{n}$, with different material-dependent elastic constants accounting for typical splay, twist and bend deformations \cite{de1993physics}.
The one-constant approximation of the OF free energy, based on the assumption of equal elastic constants, is the well-known Dirichlet energy,
\begin{equation}\label{eq:ofmodel}
  E_{\mathrm{OF}}(\mathbf{n})=K_1\int_V |\nabla \mathbf{n}|^2 \mathrm{d}\bm{x}, \quad \mathbf{n}\in \mathcal{H}^1(V,\mathbb{S}^{2}).
\end{equation}
 The defects have a straightforward interpretation in the OF framework --- they are simply discontinuities in $\mathbf{n}$ \cite{kinderlehrer_lin}. The OF theory is limited, in the sense that it cannot account for non-orientable structures, cannot account for high-dimensional line or surface defects and cannot describe biaxiality. More precisely, the OF theory cannot describe point defects in 2D and line defects in 3D, since the OF energy diverges for such defects, and both types of defects are physically observed. 

To deal with the energy blow-up issues near defects, the EL theory incorporates a scalar order parameter $s$ to model the degree of orientational order\cite{ericksen1991liquid}. The order parameter tames the singularities/defects and the EL theory with the order parameter pair, $\left(s, \mathbf{n} \right)$ can indeed describe high-dimensional line and surface defects. The zero set of $s$ is the defect set in the EL setting. A typical EL free energy is,
\begin{equation}\label{eq:elmodel}
  E_{\mathrm{EL}}(s, \mathbf{n})=\int_V \left[K|\nabla s|^{2} + s^{2} |\nabla \mathbf{n}|^{2} + \psi(s)\right] \mathrm{d}\bm{x},
\end{equation}
where $\psi$ is a bulk potential that enforces a preferred value of $s$ (typically non-zero) in the bulk.

We model NLCs within a shallow square well, $V=(-\kappa, \kappa)^2\times(0,w \kappa)$ where $ w\ll 1$. In the thin film limit, $\mathbf{n}= \left(n_1, n_2, n_2 \right)$ satisfies $\partial_{z} \mathbf{n}=\mathbf{0}$ and $n_{3}=0$, so we have $\mathbf{n}=(n_1, n_2)^\top \in \mathbb{S}^1$ in the $xy$-plane, describing the locally-preferred in-plane direction of NLC alignment \cite{golovaty2017dimension}.
Furthermore, we adopt a simple bulk energy $\psi(s)=\eta_{0} (s^{2}-s_{+}^2)^{2}$, for some given non-zero $s_+$ typically dependent on the temperature, and set $K=1$ in \eqref{eq:elmodel}.
Then the simplified Ericksen--Leslie (sEL) free energy is,
\begin{equation}\label{eq:sEL}
  E_{\mathrm{sEL}}(s, \mathbf{n})=\int_{\Omega} \left(|\nabla s|^{2} + s^{2} |\nabla \mathbf{n}|^{2} + \eta_{0}\kappa^2 (s^{2}-s_{+}^2)^{2}\right) \mathrm{d}\bm{x},
\end{equation}
where $\Omega = (-1,1)^2$ is the rescaled square domain, and $\kappa$ is the square edge length, $\eta_0$ is a material-dependent constant.
In \cite{Luigi1991A}, the authors imposed Dirichlet boundary conditions for both $s$ and $\mathbf{n}$ on two parallel plates.
In this manuscript, we impose planar degenerate boundary conditions on \cref{eq:GLmodel_EL},
\begin{equation}\label{eq:ELbc}
  \mathbf{n}\cdot\bm{\nu} = 0, \quad \bm{x}\in \partial \Omega,
\end{equation}
where $\bm{\nu}=(\nu_1,\nu_2)^\top$ is the outward unit normal vector on $\partial \Omega$, 
which can also be written in $\mathbf{p} = (p_1,p_2) = s\mathbf{n}$ as
\begin{align}
p_1 = 0\ on\ x = \pm 1,\\
p_2 = 0\ on\ y = \pm 1.
\end{align}
In particular, $\mathbf{p} \equiv \left(0,0\right)$ is a solution of this boundary-value problem which represents the \emph{isotropic} solution. Additionally, we also recover ordered solutions, such as the $D$ and $R$ solutions, as shown in subsequent sections.
The same boundary condition has also been applied in \cite{2002Axisymmetric} to discuss bipolar configurations of a NLC droplet in the OF model.

\subsection{Landau--de Gennes model}
The LdG theory is a powerful continuum theory with the $\mathbf{Q}$-tensor order parameter, a symmetric traceless $3\times 3$ matrix \cite{de1993physics}.
Let $s_{1}\geqslant s_{2} \geqslant s_{3}$ be the three eigenvalues of $\mathbf{Q}$; then we can write 
\begin{equation}\label{eq:Qn}
  \mathbf{Q}=s_{1} \mathbf{n}_{1}\otimes \mathbf{n}_{1} + s_{2} \mathbf{n}_{2}\otimes \mathbf{n}_{2} + s_{3} \mathbf{n}_{3}\otimes \mathbf{n}_{3},
\end{equation}
where $\mathbf{n}_{1}, \mathbf{n}_{2}, \mathbf{n}_{3}$ are the corresponding unit eigenvectors.
The $\mathbf{Q}$-tensor is \emph{isotropic} if $\mathbf{Q}=\mathbf{0}$, \emph{uniaxial} if $\mathbf{Q}$ has a pair of degenerate nonzero eigenvalues, and \emph{biaxial} if $\mathbf{Q}$ has three distinct eigenvalues.

In the absence of surface energies, a particularly simple form of the LdG free-energy functional is given by
\begin{equation}\label{eq:ldgenergy}
  E_{\mathrm{LdG}}(\mathbf{Q}) =
  \int_{V} \left[\dfrac{L}{2}|\nabla\mathbf{Q}|^{2}
  + \dfrac{A}{2}|\mathbf{Q}|^{2} - \dfrac{B}{3} \mathrm{tr}\; \mathbf{Q}^{3} + \dfrac{C}{4} |\mathbf{Q}|^4 \right] \mathrm{d}\bm{x},
\end{equation}
where $|\mathbf{Q}|=\sqrt{\mathrm{tr}(\mathbf{Q}^{\top}\mathbf{Q})}$ denotes the Frobenius norm of $\mathbf{Q}$, and $|\nabla\mathbf{Q}|^{2}={Q}_{ij,k}{Q}_{ij,k}$. Here, $L$ is a material-dependent elastic constant.
The polynomial terms in $\mathbf{Q}$ in \eqref{eq:ldgenergy} constitute the bulk energy potential that drives the isotropic-nematic phase transition.
For $A<0$ (where $A$ is the rescaled temperature), the bulk energy favours an ordered bulk uniaxial phase so that $\mathbf{Q} = s_+ (\mathbf{n}\otimes \mathbf{n}-\mathbf{I}/3)$ for arbitrary $\mathbf{n}\in \mathbb{S}^2$ and  $s_+ = \frac{1}{4C}(B + \sqrt{B^2 + 24|A| C})$, is the set of bulk energy minimizers for fixed material-dependent positive constants $B$ and $C$. 
In what follows, we take  $A=-B^{2}/3C$, $B=0.64\times 10^{4} \mathrm{Nm}^{-2}$, $C=0.35\times 10^{4} \mathrm{Nm}^{-2}$, and $L=4\times 10^{-11}\mathrm{N}$, which roughly correspond to the prototypical NLC material MBBA, at a characteristic low temperature \cite{Introduction}. This specific choice of parameters also aids comparison with previous work in \cite{wang2019order}, \cite{han2020pol} and \cite{han2020SL}.

In what follows, we adopt the reduced rLdG model which can be rigorously justified in the thin film limit, or in the $h \to 0$ limit \cite{sternberggolovaty2015}.
When dealing with 2D systems, the reduced Landau--de Gennes (rLdG) model has been successfully applied, both for capturing the qualitative properties of physically relevant solutions and for probing into defect cores \cite{brodin2010melting, bisht_epl}.
Except for the phenomena referred to as ``escape into the third dimension'' \cite{sonnet1995alignment}, the physically relevant $\mathbf{Q}$-tensors on 2D domains have a fixed eigenvector $\mathbf{z}$, the unit vector in the $z$-direction, and can be written as,
\begin{equation}\label{eq:Qred}
\mathbf{Q} = q_1\left( \mathbf{n}\otimes \mathbf{n} - \mathbf{m}\otimes \mathbf{m} \right) + q_2 \left(\mathbf{n}\otimes \mathbf{m} + \mathbf{m}\otimes \mathbf{n} \right)+ q_3 \left( 2\mathbf{z}\otimes \mathbf{z} - \mathbf{n}\otimes \mathbf{n} - \mathbf{m}\otimes \mathbf{m} \right).\nonumber
\end{equation}
Here $\mathbf{n}$ and $\mathbf{m}$ are orthonormal vectors in the $xy$-plane \cite{sternberggolovaty2015}, with only three degrees of freedom $q_1,q_2,q_3$ (out of five) in a 2D framework.
Furthermore, for $A = -\frac{B^2}{3C}$, $q_3$ is constant for all physically relevant critical points of the form \eqref{eq:Qred}. 
Hence, for $A=-\frac{B^2}{3C}$, we have a reduced description in terms of a reduced tensor, $\mathbf{Q}_r$, with only two degrees of freedom such that
\begin{equation}\label{eq:QP}
\mathbf{Q} =
\left(\begin{tabular}{cc|c}
\multicolumn{2}{c|}{\multirow{2}*{$\mathbf{Q}_r+\dfrac{s_+}{6}\mathbf{I}_2$}} & $0$ \\
\multicolumn{2}{c|}{} & $0$ \\ \hline
$0$ & $0$ & $-\dfrac{s_+}{3}$ \\
\end{tabular}\right),
\end{equation}
where $\mathbf{I}_2$ is the $2\times 2$ identity matrix.
The reduced LdG order parameter 
\begin{equation}
\mathbf{Q}_r = \left(\begin{tabular}{cc}
$q_1$ & $q_2$ \\
$q_2$ & $-q_1$ \\
\end{tabular}\right)
\end{equation}
is a $2\times 2$ symmetric traceless tensor, with pnly two independent components.

The zero set of $\mathbf{Q}_r$ is the set of uniaxial $\mathbf{Q}$-tensors with the negative order parameter about $\mathbf{z}$. The zero set of $\mathbf{Q}_r$ describes the planar defects on the square domain.
By substituting the relation between $\mathbf{Q}$ and the reduced tensor, $\mathbf{Q}_r$, in \eqref{eq:QP}, we obtain the rLdG energy functional to be
\begin{equation}\label{eq:rldgenergy2d}
  E_{\mathrm{rLdG}}(\mathbf{Q}_r)=\int_{\Omega} \left[\dfrac {1}{2}|\nabla \mathbf{Q}_r|^2 +\frac{\alpha}{8} \left(|\mathbf{Q}_r|^2-\dfrac{s_+^2}{2}\right)^2\right]\mathrm{d}\bm{x}.
\end{equation}
$\Omega=(-1,1)^2$ is a rescaled unit square and $\alpha=2\kappa^2C/L>0$ describes the square size. 

We impose planar Dirichlet boundary conditions on $\partial\Omega$,
\begin{equation}\label{eq:ldgbc}
  \mathbf{Q}_r|_{\partial \Omega} = \mathbf{Q}_{r\mathrm{b}}=\left\{
\begin{split}
\left(
\begin{tabular}{cc}
$\frac{B}{2C}$ & $0$ \\
$0$ & $-\frac{B}{2C}$ \\
\end{tabular}\right),\;&x\in(-1 + \sigma,1-\sigma),\ y=\pm1,\\
\left(
\begin{tabular}{cc}
$-\frac{B}{2C}$ & $0$ \\
$0$ & $\frac{B}{2C}$ \\
\end{tabular}\right),\;&x=\pm1,\ y\in(-1 + \sigma,1-\sigma),
\end{split}\right.
\end{equation}
which can also be written in terms of $\mathbf{q}=(q_1,q_2)^\top$ as,
\begin{equation}
  \mathbf{q}|_{\partial \Omega} =\mathbf{q}_{\mathrm{b}}=
  \begin{cases}
\frac{B}{2C}(1,0) & x\in(-1 + \sigma,1-\sigma),\ y=\pm1,\\
\frac{B}{2C}(-1,0) & x=\pm1,\ y\in(-1 + \sigma,1-\sigma).
  \end{cases}
\end{equation}
where $0<\sigma \ll 1$ is the size of mismatch region. We set the value at the vertices to be $q_1 = 0$ and $q_2 = 0$. On the $\sigma$-neighbourhood of the vertices, we linearly interpolate between the constant values in \eqref{eq:ldgbc} and the average value at the vertex. For $\sigma$  sufficiently small, this interpolation does not affect solution properties in the interior.

\subsection{The relationship between the sEL and rLdG models}
The GL theory  is predominantly used to study the phenomena of superconductivity in \cite{ginzburg2009theory}.
A simple GL energy functional can be written as,
\begin{equation}\label{eq:GL}
E_{\mathrm{GL}}(\mathbf{u}) = \int_{\Omega} \left[\frac{1}{2}|\nabla \mathbf{u}|^2 + \frac{1}{4\epsilon_u^2}\left(|\mathbf{u}|^2-1\right)^2\right] \mathrm{d}\bm{x},
\end{equation}
where $\mathbf{u}\in \mathcal{H}^1(\Omega;\mathbb{R}^2)$ and $\epsilon_u$ is a material-dependent constant.
It is a prototype model in materials science, extensively studied by many researchers.
The GL functional is nonlinear, with typically multiple critical points, and these critical points can be classified in terms of the topological degrees and locations of the point defects \cite{bethuel1994ginzburg}.

In the sEL model, we employ the re-scalings: $\tilde{s} = s/s_+$, and substitute $\mathbf{p}=s\mathbf{n}=(p_1, p_2)^\top$ into \eqref{eq:sEL}, so that the sEL free energy \eqref{eq:sEL} reduces to the GL free energy,
\begin{equation}\label{eq:GLmodel_EL}
  E_{\mathrm{sEL}}(\mathbf{p})=\int_{\Omega}
  \left[\dfrac{1}{2}|\nabla \mathbf{p}|^{2} + \dfrac{\eta}{4}\left(|\mathbf{p}|^{2}-1\right)^{2}\right]\mathrm{d}\bm{x},
\end{equation}
where $\eta=2\eta_{0}s_{+}^{2}\kappa^{2}$ encodes information about the domain size i.e $\kappa^2$ is the area of the square domain.

The Euler--Lagrange equations of the free energy \cref{eq:GLmodel_EL}, with the boundary conditions in \cref{eq:ELbc}, are given by,
\begin{equation}\label{eq:eleq}
\left\{
\begin{aligned}
  \Delta \mathbf{p} = \eta(|\mathbf{p}|^{2}-1)\mathbf{p},\qquad&\bm{x} \in \Omega,\\
  \mathbf{p}\cdot\bm{\nu} = \nabla\times\mathbf{p} = 0,\qquad&\bm{x} \in \partial\Omega,\\
\end{aligned}
\right.,
\end{equation}
with a trivial solution  $\mathbf{p}(\bm{x})\equiv(0,0)^{\top}$ (a homogeneous isotropic state), for all $\eta>0$.
Furthermore, for $\eta>0$ small enough, the isotropic state is the unique solution of \eqref{eq:eleq}, and the global minimizer of \eqref{eq:GLmodel_EL} \cite{Lamy2014Bifurcation}.

The boundary condition, $\nabla\times\mathbf{p} = 0$, is derived from variational arguments as follows.
For $\mathbf{p},\mathbf{u} \in \mathcal{A} = \{\mathbf{p}\in\mathcal{H}^1(\Omega, \mathbb{R}^2): \mathbf{p}|_{\partial\Omega}\cdot\bm{\nu} = 0\}$, where $\bm\nu$ is piecewise constant on $\partial \Omega$ (the square edges), we have,
\begin{align}
&\langle\nabla E_{\mathrm{sEL}}(\mathbf{p}),\mathbf{u}\rangle =
\int_{\Omega} \left(\nabla \mathbf{p}: \nabla \mathbf{u}+  \eta(|\mathbf{p}|^2-1)\mathbf{p}\cdot\mathbf{u} \right)\mathrm{d}\bm{x}\nonumber\\
& = \int_{\Omega} \left(-\Delta\mathbf{p}+\eta(|\mathbf{p}|^2-1)\mathbf{p}\right)\cdot\mathbf{u}\mathrm{d}\bm{x}
+ \int_{\partial\Omega} \mathbf{u} \cdot\partial_{\bm \nu}\mathbf{p}\mathrm{d}\sigma 
\nonumber\\
& = \int_{\Omega} \left(-\Delta\mathbf{p}+\eta(|\mathbf{p}|^2-1)\mathbf{p}\right)\cdot\mathbf{u}\mathrm{d}\bm{x} + \int_{\partial\Omega} \left(
(\bm{\nu} \times \mathbf{u}) (\nabla\times \mathbf{p}) 
+ (\mathbf{u}\cdot \nabla) (\mathbf{p} \cdot \bm{\nu})
\right)\mathrm{d}\sigma.
\end{align}
Since $\mathbf{p} \cdot \bm{\nu}=0$ and $\mathbf{u}\cdot \bm{\nu} =0$ hold on $\partial \Omega$, the critical points $\mathbf{p}$ satisfy $\nabla\times\mathbf{p} = 0$.

In the rLdG model, with the two-dimensional vector field $\mathbf{q} = (q_1,q_2)^\top$, the rLdG free energy \eqref{eq:rldgenergy2d} can be rewritten as a GL-type free-energy,
\begin{equation}\label{eq:GLmodel_LdG} 
E_{\mathrm{rLdG}}(\mathbf{q}) = 2\int_{\Omega} \left[\dfrac {1}{2}|\nabla \mathbf{q}|^2 +\frac{\alpha}{4} \left(|\mathbf{q}|^2-\dfrac{s_+^2}{4}\right)^2\right]\mathrm{d}\bm{x}.
\end{equation}
Substituting $\tilde{\mathbf{q}} = 2\mathbf{q}/s_+$,
the rLdG free energy \eqref{eq:GLmodel_LdG} reduces to the GL energy (tildes omitted),
\begin{equation}\label{eq:GLmodel_LdG1}
  E_{\mathrm{rLdG}}(\mathbf{q})= \frac{s_+^2}{2} \int_{\Omega} \left[\dfrac {1}{2}|\nabla \mathbf{q}|^2 +\frac{\alpha s_+^2}{16} \left(|\mathbf{q}|^2-1\right)^2\right]\mathrm{d}\bm{x},
\end{equation}
where $\alpha$ is an area parameter that will be motivated in the next subsection.
The corresponding Euler--Lagrange equations are 
\begin{equation}\label{eq:rLdGeleq}
\Delta \mathbf{q} = \frac{\alpha s_+^2}{4}\left(|\mathbf{q}|^2-1\right)\mathbf{q},
\end{equation}
with Dirichlet boundary conditions \eqref{eq:ldgbc} on $\partial\Omega$.

In the sEL model, we have two order parameters, the nematic director $\mathbf{n}$, and the nematic order $s$. 
In rLdG model, the reduced tensor $\mathbf{Q}_r$ can also be expressed in terms of $s$ and $\mathbf{n}$ as,
\begin{equation}\label{eq:relation_LdG}
\mathbf{Q}_r = s(\mathbf{n}\otimes\mathbf{n}-\mathbf{I}/2).
\end{equation}
Regarding the boundary conditions \eqref{eq:ELbc} for the sEL model and \eqref{eq:ldgbc} for the rLdG model, the director $\mathbf{n}$ is constrained to be tangent to the square edges in both cases but $s$ is free in \eqref{eq:ELbc}, and fixed to be $s = \frac{B}{C}$ in \eqref{eq:ldgbc}.
In other words, the rLdG model has a more strongly-anchored boundary condition, which naturally enforces strongly ordered regions near the square edges as we will see below.

\subsection{Matching the domain sizes of the sEL and rLdG models}\label{matching_size}
For the purposes of comparisons between the sEL and rLdG models, we match the effective domain sizes in \eqref{eq:GLmodel_EL} and \eqref{eq:GLmodel_LdG1} by considering the parameters in the associated GL-type functionals. Comparing the GL-type functionals \eqref{eq:GLmodel_EL} with \eqref{eq:GLmodel_LdG1}, we deduce that if the effects of the rLdG and the sEL boundary conditions are relaxed on the edges of a square domain $\Omega$, the effective dimensionless area parameter $\eta_e$ in the sEL model corresponds to  $\alpha_e = \frac{4\eta_e}{s_+^2}\approx 1.2\eta_e$ (for $A = -\frac{B}{3C}$) in the rLdG model.

The effect of the different boundary conditions in \eqref{eq:ELbc} and \eqref{eq:ldgbc} should also be taken into consideration on the domain sizes.
There is no constraint for nematic order $s$ on the boundary condition in the sEL model, so all the domain can be effective to accommodate defects, i.e. $\eta = \eta_e$.
The Dirichlet boundary condition \eqref{eq:ldgbc} for the rLdG model has a stronger impact on the degree of order near the square edges.
For example, for the $BD$ solutions in the rLdG setting (Figure \ref{fig:1}), there are two line defects near a pair of opposite square edges, such that $\mathbf{n}$ is almost uniform between each line defect (with $\mathbf{q}\approx \mathbf{0}$) and the adjacent square edge. 
At a given value of $\alpha$, we define $\alpha_b(\alpha)$ to be the area of each region of almost uniform alignment of $\mathbf{n}$, near one square edge in the $BD$ solutions. In practice, $\alpha_b(\alpha)$ is calculated as $\frac{\alpha}{4}n_b h^2$, where $h$ is the mesh size, $n_b(\alpha)$ is the number of mesh elements between the line defect and the adjacent square edge in the $BD$ solution on $\Omega$ at $\alpha$, and $\alpha/4$ the scaling ratio between the area measure $\alpha$ and the area of $\Omega$.
Hence, the renormalized dimensionless area parameter $\alpha$ should be $\alpha = \alpha_e+4\times\alpha_b$.

The matching $\eta$ in the sEL model and $\alpha$ in the rLdG model are listed in Table \ref{table1}. 
\begin{table}[htbp]
	\centering
	\begin{tabular}{ccc|cccccccccc}
		\hline
		&$\eta$&&&1&&3&&6&&12&&20\\  
		\hline
		&$\alpha$&&&5&&15&&18&&30&&47\\
		\hline
	\end{tabular}
	\caption{The domain size $\eta$ in the sEL model and matched domain size $\alpha$ in the rLdG model.}\label{table1}
\end{table}

\section{Numerical algorithm of the solution landscape with symmetry properties} \label{sec:algorithm_construction}
The physical systems often have some intrinsic symmetry properties that can be ultilized to reduce computational costs.
Assume that $\Gamma$ is the symmetry group of the energy functional $E$ on a linear manifold $\mathcal{M}$ embedded in the Hilbert space $\mathcal{H}$, that is, $\forall \mathcal{T}\in \Gamma$ is an isometric operator on $\mathcal{M}$ satisfying,
\begin{equation}\label{eq:symene1}
  E(\mathbf{r})=E(\mathcal{T}\mathbf{r}), \forall \mathbf{r}\in \mathcal{M}.
\end{equation}
Then we have,
\begin{equation}
\begin{aligned}\label{eq:symene2}
\nabla E(\mathcal{T}\mathbf{r})&=\mathcal{T}\nabla E(\mathbf{r}), \forall \mathbf{r}\in \mathcal{M},\\
\langle \mathcal{T} \mathbf{v},\nabla^2 E(\mathcal{T}\mathbf{r})\mathcal{T}\mathbf{v}\rangle&=\langle\mathbf{v},\nabla^2 E(\mathbf{r})\mathbf{v}\rangle, \forall \mathbf{r}\in \mathcal{M}, \forall\mathbf{v}\in \mathcal{M}_0,
\end{aligned}
\end{equation}
where $\mathcal{M}_0=\{\mathbf{r}_1-\mathbf{r}_2: \mathbf{r}_1, \mathbf{r}_2\in \mathcal{M}\}$ is a subspace of $\mathcal{H}$.
Although the energy is invariant under the symmetry group $\Gamma$, the critical points may be invariant only under a subgroup of $\Gamma$ \cite{troger2012nonlinear}.
Therefore, once we obtain a critical point $\mathbf{p}$, we may obtain plenty of critical points $\Gamma\mathbf{p}$ with the same energy and index.
We present a specific example, the NLCs in a two-dimensional regular polygon domain with $K$ edges, to illustrate the symmetry properties.
For a domain size large enough, we have at least $\lfloor K/2\rfloor$ classes of stable states and $K\choose 2$ stable states \cite{han2020pol}. 
In a square domain, there are two $D$ solutions for which the nematic alignments are almost parallel to one diagonal and four $R$ solutions for which the director rotates by $\pi$ radians between opposite edges \cite{kusumaatmaja2015free}.
A disc domain can be viewed as a limit situation of $K\to\infty$, and there is only one type of solution, the Planar Polar state, with an infinite number, featured by two interior point defects along a disc diameter \cite{han2019transition}.
Hence, the study of symmetry properties is essential for considering critical points.

In the sEL model, the symmetry group is,
\begin{equation}
\Gamma_{\mathrm{sEL}}=\langle \mathcal{R}, \mathcal{S}, \mathcal{T}_1, \mathcal{T}_2 \rangle,
\end{equation}
with $32$ elements, where the operators are defined as,
\begin{equation*}\label{eq:GLsym1}
\begin{aligned}
&\mathcal{R}\mathbf{p}(\bm{x})=
\begin{bmatrix}-p_{2} \\ p_{1} \\ \end{bmatrix}(-y, x),&
&\mathcal{S}\mathbf{p}(\bm{x})=
\begin{bmatrix}-p_{1} \\ p_{2} \\ \end{bmatrix}(-x, y),&\\
&\mathcal{T}_{1}\mathbf{p}(\bm{x})=
\begin{bmatrix}-p_{1} \\ p_{2} \\ \end{bmatrix}(x,y),&
&\mathcal{T}_{2}\mathbf{p}(\bm{x})=
\begin{bmatrix} p_{1} \\-p_{2} \\ \end{bmatrix}(x,y).&
\end{aligned}
\end{equation*}
For a critical point, using symmetry properties, we can get at most $31$ additional critical points with the same energy and Morse index.

In the rLdG model, the energy functional of the rLdG model \eqref{eq:rldgenergy2d} is defined on the linear manifold,
\begin{gather}\label{eq:ldgdef}
  \mathcal{M}=\{\mathbf{Q}_r\in \mathcal{H}^{1}(\Omega,\mathbb{R}^{2\times2}):  \mathbf{Q}_r=\mathbf{Q}_r^\top, 
   \mathrm{tr} \;\mathbf{Q}_r=0,
   \mathbf{Q}|_{\partial \Omega} = \mathbf{Q}_{r\mathrm{b}}, \},
\end{gather}
and the symmetry group is
\begin{equation}
\Gamma_{\mathrm{rLdG}}=\langle \mathcal{R}, \mathcal{S}, \mathcal{T}\rangle,
\end{equation}
with $16$ elements defined as
\begin{eqnarray*}\label{eq:GLsym2}
\mathcal{R} \mathbf{Q}_r(\bm x)=\begin{bmatrix}
-q_1 & q_2 \\ q_2 & q_1	\\ \end{bmatrix}(-y, x),
\mathcal{S} \mathbf{Q}_r(\bm x)=\begin{bmatrix}
q_1 & q_2\\q_2 & -q_1\\\end{bmatrix}(-x, y),
\mathcal{T} \mathbf{Q}_r(\bm x)=\begin{bmatrix}
q_1   & -q_2 \\-q_2	& -q_1 \\\end{bmatrix}(x, y).
\end{eqnarray*}

The solution landscapes are constructed by the downward/upward search algorithms combined with the SD method for finding all connected saddle points and minima \cite{yin2020construction}.
By taking symmetry properties into consideration, we propose a simplified algorithm for the solution landscape and the computational costs are greatly reduced by leaving out unnecessary calculations for critical points.
Specifically, the solution landscape is constructed by two algorithms where the symmetry properties are considered,
a downward search algorithm that enables us to search for all connected lower-index saddles from a given index-$k$ saddle, and
an upward search algorithm with a selected direction to find the higher-index saddles.
As long as the symmetry properties have been captured, this algorithm can be applied on the energy functional $E$.

{\it Downward search.} Given an index-$k$ saddle point $\hat{ \mathbf{r}}$ and $k$ unit orthonormal eigenvectors $\hat{ \mathbf{v}}_1,\cdots,\hat{ \mathbf{v}}_k$ corresponding to the smallest $k$ negative eigenvalues $\hat{\lambda}_1\leqslant \cdots\leqslant\hat{\lambda}_k$ of the Hessian $\nabla^2 E( \hat{\mathbf{r}})$.
We search for an index-$m$ ($m<k$) saddle point by applying the SD method for an index-$m$ saddle ($m$-SD) \cite{yin2019high},
\begin{equation}\label{eq:dynamics}
\left\{
\begin{aligned}
\dot{ \mathbf{r}}   & =- \left(\bm I-2\sum_{j=1}^{m} \mathbf{v}_j \mathbf{v}_j^\top\right)\nabla E( \mathbf{r}), \\
\dot{ \mathbf{v}}_i & = -\left(\bm I-\mathbf{v}_i\mathbf{v}_i^\top-2\sum_{j=1}^{i-1} \mathbf{v}_j \mathbf{v}_j^\top\right)\nabla^2 E(\mathbf{r}) \mathbf{v}_i, i=1,\cdots,m,\\
\end{aligned}
\right.
\end{equation}
where $\bm I$ is the identity operator.
To avoid the explicit calculation of Hessian matrix, $\nabla^2 E(\mathbf{r})\mathbf{v}_i$ can be approximated by
\begin{equation}
\nabla^2 E(\mathbf{r}) \mathbf{v}_i \approx \dfrac{\nabla E( \mathbf{r}+l \mathbf{v}_i) - \nabla E( \mathbf{r}-l \mathbf{v}_i)}{2 l}.
\end{equation}
We start the $m$-SD method from an initial point $ \mathbf{r}(0) = \hat{ \mathbf{r}}\pm\varepsilon \hat{\mathbf{v}}_j$ where $\hat{\mathbf{v}}_j$ is chosen from the unstable eigenvectors, and $m$ initial unstable directions $\mathbf{v}_i(0)$ are chosen from the unstable directions except $\hat{\mathbf{v}}_{j}$.
If the dynamics converges to a new solution $\tilde{ \mathbf{r}}$, the connection $( \hat{ \mathbf{r}},\tilde{ \mathbf{r}})$ is also obtained in the solution landscape.
From the symmetry group, $\forall \mathcal{T}\in \Gamma$, $\mathcal{T}\tilde{ \mathbf{r}}$ is also a critical point with the same energy and Morse index, which can be found following a similar $m$-SD from $\mathcal{T} \hat{ \mathbf{r}}$, indicating the connection $(\mathcal{T} \hat{ \mathbf{r}},\mathcal{T}\tilde{ \mathbf{r}})$.
The downward search with the symmetry group $\Gamma$ is shown in Algorithm \ref{alg:Dsymmetrygroup}.

\begin{algorithm}[H]
\caption{Downward search with a symmetry group $\Gamma$}
\label{alg:Dsymmetrygroup}
\begin{algorithmic}[1]
\REQUIRE{A $k$-saddle $\hat{ \mathbf{r}}$, $\varepsilon>0$.}
\STATE{Calculate the $\hat{k}$ eigenvectors $\hat{ \mathbf{v}}_1,\cdots,\hat{ \mathbf{v}}_{k}$ of $\nabla^2 E(\hat{ \mathbf{r}})$;}
\STATE{Set the queue $\mathcal{A}= \{(\hat{ \mathbf{r}},k-1,\{\hat{ \mathbf{v}}_1,\cdots,\hat{ \mathbf{v}}_{k}\})\}$,
the solution set $\mathcal{Z}=\{\hat{\mathbf{r}}\}$, and the relation set $\mathcal{C}=\varnothing$;}
\WHILE{$\mathcal{A}$ is not empty}
  \STATE{Pop $(\mathbf{r},m,\{\mathbf{v}_1,\cdots,\mathbf{v}_n\})$ from $\mathcal{A}$;}
  \STATE{Push $(\mathbf{r},m-1,\{\mathbf{v}_1,\cdots,\mathbf{v}_n\})$ into $\mathcal{A}$ if $m\geqslant1$;}
  \FOR{$j=1:k$}
   \STATE{Determine the initial directions: $\{\mathbf{v}_i: i=1, \cdots, m+1, i\neq \min(j,m+1)\}$;}
    \IF{$m$-SD from $\mathbf{r}\pm \varepsilon \mathbf{v}_j$ converges to $(\tilde{\mathbf{r}}, \tilde{\mathbf{v}}_1, \cdots, \tilde{\mathbf{v}}_m)$}
      \STATE{$\mathcal{C}\leftarrow\mathcal{C} \cup \{(\mathbf{r},\tilde{\mathbf{r}})\}$;}
      \IF{$\Gamma\tilde{\mathbf{r}} \cap \mathcal{S}=\varnothing$}
        \STATE{$\mathcal{Z}\leftarrow\mathcal{Z} \cup \{\tilde{\mathbf{r}}\}$;}
        \STATE{Push $(\tilde{\mathbf{r}},m-1,\{\tilde{\mathbf{v}}_1,\cdots,\tilde{\mathbf{v}}_{m}\})$ into $\mathcal{A}$ if $m \geqslant 1$;}
      \ENDIF
    \ENDIF
  \ENDFOR
\ENDWHILE
\ENSURE{The solution set $\mathcal{Z}$ and the relation set $\mathcal{C}$.}
\end{algorithmic}
\end{algorithm}

{\it Upward search.}
Given an index-$k$ saddle $\hat{\mathbf{r}}$, to search for an index-$K$ ($K>k$) saddle,
the eigenvectors $\hat{\mathbf{v}}_1,\cdots,\hat{\mathbf{v}}_K$ corresponding to the smallest $K$ eigenvalues are calculated, where $\hat{\mathbf{v}}_i$ ($i\leqslant k$) is an unstable direction, and $\hat{\mathbf{v}}_i$ ($i> k$) is a stable direction.
The initial state $\mathbf{r}(0)$ is set as $\hat{\mathbf{r}}\pm\varepsilon \hat{\mathbf{v}}_j$ where $\hat{\mathbf{v}}_j$ is chosen from the stable directions $\{\hat{\mathbf{v}}_{k+1},\cdots,\hat{\mathbf{v}}_K\}$, and the $m$ initial directions $\mathbf{v}_i(0)$ should include $\hat{\mathbf{v}}_j$.
A typical initial condition for $m$-SD in an upward search is $(\hat{\mathbf{r}}\pm\varepsilon \hat{\mathbf{v}}_{m}, \hat{\mathbf{v}}_{1},\cdots, \hat{\mathbf{v}}_m)$.
If the dynamics converges to a new solution $\tilde{\mathbf{r}}$, $\Gamma\tilde{\mathbf{r}}$ are also solutions with the same energy and index. 
The upward search with the symmetry group $\Gamma$ is presented with the typical choice in Algorithm \ref{alg:Dsymmetrygroup}.

\begin{algorithm}[H]
\caption{Upward search with the symmetry group $\Gamma$}
\label{alg:Usymmetrygroup}
\begin{algorithmic}[1]
\REQUIRE{A $k$-saddle $\hat{\mathbf{r}}$, $\varepsilon>0$, the highest searching index $K$.}
\STATE{Calculate the $K$ eigenvectors $\hat{\mathbf{v}}_1,\cdots,\hat{\mathbf{v}}_{K}$ of $\nabla^2 E(\hat{\mathbf{r}})$;}
\STATE{Set the stack $\mathcal{A}= \{(\hat{\mathbf{r}},k+1,\{\hat{\mathbf{v}}_1,\cdots,\hat{\mathbf{v}}_{K}\})\}$,
the solution set $\mathcal{Z}=\{\hat{\mathbf{r}}\}$;}
\WHILE{$\mathcal{A}$ is not empty}
  \STATE{Pop $(\mathbf{r},m,\{\mathbf{v}_1,\cdots,\mathbf{v}_K\})$ from $\mathcal{A}$;}
  \STATE{Push $(\mathbf{r},m+1,\{\mathbf{v}_1,\cdots,\mathbf{v}_K\})$ into $\mathcal{A}$ if $m\geqslant1$;}
    \IF{$m$-SD from $\mathbf{r}\pm \varepsilon \mathbf{v}_m$ converges to $(\tilde{\mathbf{r}}, \tilde{\mathbf{v}}_1, \cdots, \tilde{\mathbf{v}}_m)$}
      \IF{$\Gamma\tilde{\mathbf{r}} \cap \mathcal{Z}=\varnothing$}
        \STATE{Calculate the corresponding orthonormal eigenvectors $\{\tilde{\mathbf{v}}_1,\cdots,\tilde{\mathbf{v}}_K\}$ of $\nabla^2 E(\tilde{\mathbf{r}})$;}
        \STATE{$\mathcal{Z}\leftarrow\mathcal{Z} \cup \{\tilde{\mathbf{r}}\}$;}
        \STATE{Push $(\tilde{\mathbf{r}},m+1,\{\tilde{\mathbf{v}}_1,\cdots,\tilde{\mathbf{v}}_{K}\})$ into $\mathcal{A}$ if $m \leqslant K$;}
      \ENDIF
    \ENDIF
\ENDWHILE
\ENSURE{The solution set $\mathcal{Z}$.}
\end{algorithmic}
\end{algorithm}
The combination of these two algorithms drives the entire search to navigate up and down systematically and efficiently on the energy landscape. 
We are able to find a number of critical points and discover the connectivity of the solution landscape. 
In what follows, we apply the downward/upward search algorithms with symmetry properties to efficiently compute the critical points, including both saddle points and minima, for the sEL and rLdG models on a square domain.
The finite difference method is implemented with a mesh size $h = 1/50$ to approximate the spatial derivatives for the rescaled square domain $\Omega$.
An explicit Euler scheme with stepsize $\Delta t = 0.5h^2$ is applied in the time evolution of \eqref{eq:dynamics}.

\section{Results}\label{sec:results}
We now apply this numerical algorithm to compute the solution landscapes of the sEL model, with increasing domain size parameter $\eta$, and compare them with the solution landscapes of the rLdG model with matching domain size $\alpha$, as given in Table \ref{table1}.

\subsection{Small domain ($\eta = 1$ and $\eta = 3$)}

\begin{figure*}[htbp]
\centering
    \includegraphics[width=0.8\textwidth]{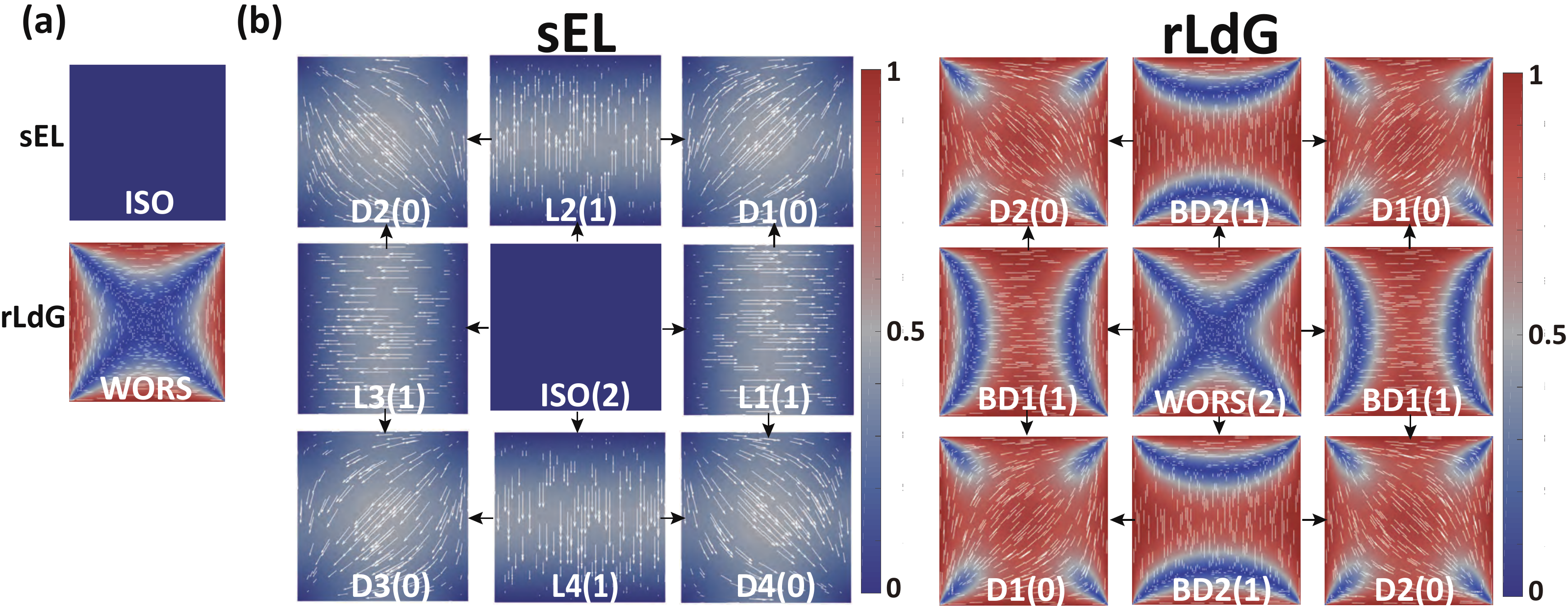}
    \caption{(a) The plot of the $ISO$ solution in sEL model at $\eta = 1$ and the $WORS$ solution at $\alpha = 5$ in rLdG model. (b) The solution landscapes of sEL model at $\eta = 3$ and rLdG model at $\alpha = 15$. The black arrow from higher-index state to lower-index state represents the connection between them. The white arrows or lines represent nematic direction $\mathbf{n}$, and the color from blue to red represents the rescaled nematic order $s$ ($|\mathbf{p}|$ in sEL model \eqref{eq:GLmodel_EL} and $|\mathbf{q}|$ in rLdG model \eqref{eq:GLmodel_LdG1}) from $0$ to $1$. All the subsequent figures of the solutions in sEL model or rLdG model have the same color bar for nematic order.}
    \label{fig:1}
\end{figure*}

The isotropic ($ISO$) state $\mathbf{p} = (0,0)^{\top}$ in the sEL model is shown in Figure \ref{fig:1}(a). For $\mathbf{p} = (0,0)^{\top}$, the bulk term $\left(|\mathbf{p}|^{2}-1\right)^{2}$ in the sEL model \eqref{eq:GLmodel_EL} attains its maximum. For $\eta = 1$, the $ISO$ state is the unique global sEL energy minimizer. As the domain size increases, i.e., the area of the isotropic phase increases and the index of the $ISO$ solution increases rapidly. 

For the rLdG model on a 2D square domain, the well order reconstruction solution ($WORS$), featured by mutually orthogonal defect lines along the two square diagonals, is shown in Figure \ref{fig:1}(a).
For the domain size $\alpha$ small enough (\emph{e.g.} $\alpha\leqslant 5$), the $WORS$ is the unique solution of \eqref{eq:rLdGeleq}, and hence, the global minimizer of the rLdG free energy \eqref{eq:rldgenergy2d} \cite{canevari2017order}. 
The $WORS$ solution exists for any domain size. 
As the edge length increases, the diagonal defect lines become longer, and its Morse index increases \cite{yin2020construction}. 
Thus, when we construct the solution landscapes for different domain sizes, the $ISO$ solution and the $WORS$ solution are always chosen to be the parent states (the saddle points with the highest index) in the sEL model and rLdG model, respectively.

For the sEL model with $\eta = 3$ (Figure \ref{fig:1}(b)), the $ISO$ state is no longer stable and is index-$2$, with a pair of degenerate eigenvalues. The index-$2$ $ISO$ state connects to four index-$1$ $L$ states ($L_1$, $L_2$, $L_3$, $L_4$). The $L$ states are featured by a constant director, $\mathbf{n}$ along the $x$ or $y$-axis ($(1,0)$, $(0,1)$, $(-1,0)$, $(0,-1)$), with two boundary line defects on the opposite edges orthogonal to the constant director. 
Each $L$ state further connects to two stable (index-$0$) $D$ states. The corresponding directors in $D$ states ($D_1$, $D_2$, $D_3$, $D_4$) are almost along the square diagonals ($(1,1)$, $(-1,1)$, $(-1,-1)$, $(1,-1)$) with four point defects at the vertices. The orientability of the sEL order parameter essentially renders two $D$ states for each square diagonal.

For the corresponding rLdG model with $\alpha = 15$ (Figure \ref{fig:1}(b)), we have the index-$2$ $WORS$, analogous to the $ISO$ state, with low order at the square center. The index-$1$ $BD$ states have two line defects near a pair of opposite square edges, with constant $\mathbf{n}$ between the line defects, analogous to the  analogue index-$1$ $L$ states. Finally, we have the stable index-$0$ $D$ states, and as in the sEL model, the $WORS$ connects to the $BD$, and the $BD$ connects to the $D$ solutions.

The sEL model does not respect the head-to-tail symmetry i.e. $\mathbf{n}$ and $-\mathbf{n}$ are distinct, in contrast to the rLdG model. Recall that $\mathbf{Q}_r= s(\mathbf{n}\otimes\mathbf{n}-\frac{\mathbf{I}}{2})$, so that $\mathbf{n}$ and $-\mathbf{n}$ correspond to the same $\mathbf{Q}_r$. Therefore, the number of $BD$ and $D$ states in rLdG model is half the number of $L$ and $D$ states in the sEL model. The repetitive images of $BD$ and $D$ in Figure \ref{fig:1}(b) are for the sake of comparison. Further, we note the much reduced order near the edges for the sEL critical points, in comparison to the rLdG case. This is simply because there is no preferred value of $s$ on $\partial \Omega$ in the sEL setting, and for $\eta$ small, there is a small energetic penalty associated with small $|\mathbf{p}|$, whereas small $|\mathbf{p}|$ reduces the elastic energy cost. In the rLdG case, $s$ is fixed on the edges and this naturally propagates into the square interior, to reduce elastic energy costs, for all values of $\alpha$.

\subsection{Moderate domain ($\eta = 6$ and $\eta = 12$)}
\begin{figure}[htbp]
    \begin{center}
    \includegraphics[width=0.5\columnwidth]{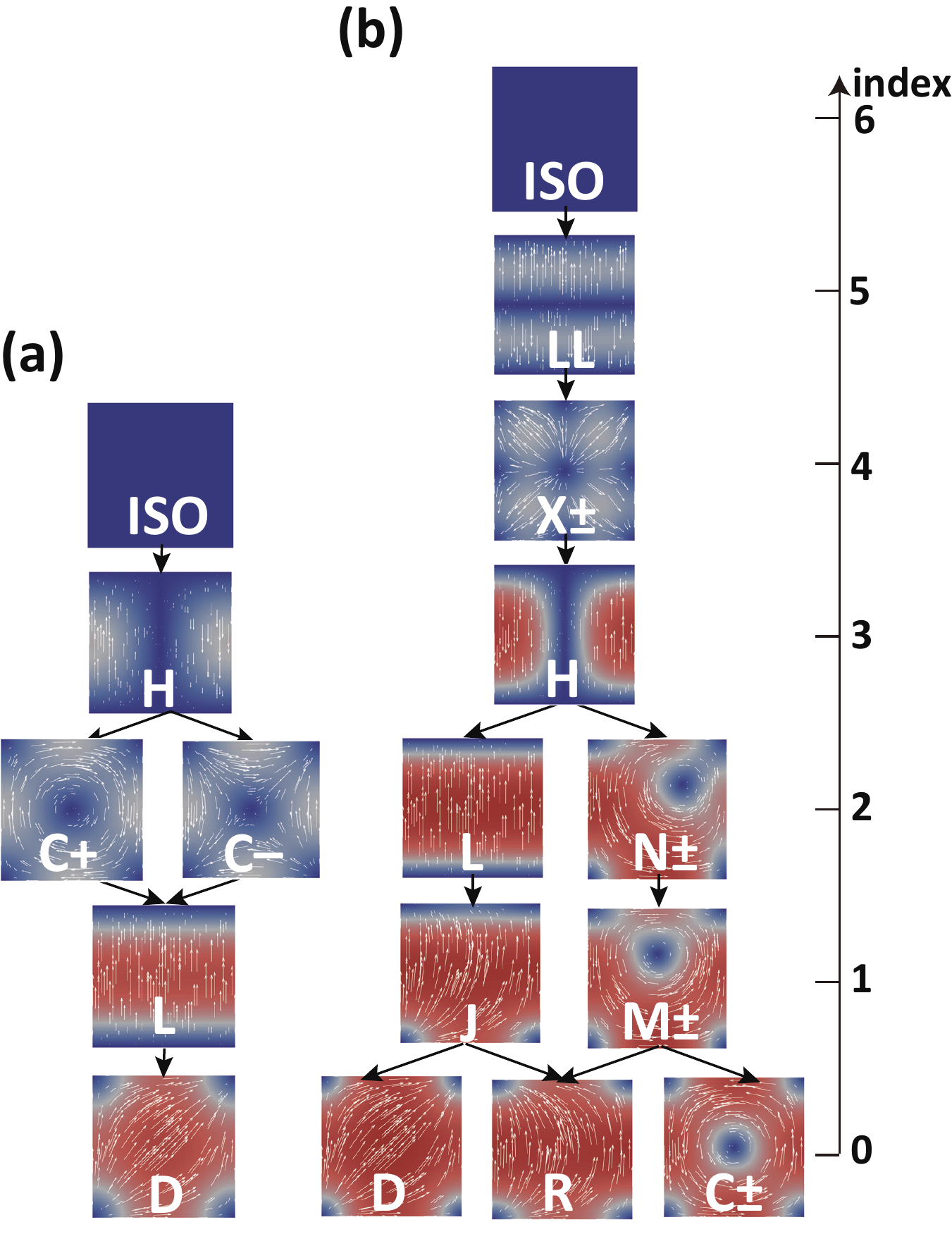}
    \caption{Solution landscapes of sEL model at $\eta = 6$ and $12$. The height of a configuration corresponds to its Morse index. The configuration of $C-$ at $\eta=12$ is analogous to $\eta = 6$, and we omit it for brevity.}
    \label{fig:2}
    \end{center}
\end{figure}

As the domain size $\eta$ increases from $3$ to $6$ in the sEL model, the index of the parent state, $ISO$, increases from $2$ to $4$. In Figure \ref{fig:2}(a), we numerically obtain new index-$3$ $H$ and index-$2$ $C\pm$ states, which are connected to the index-$4$ $ISO$ to the index-$1$ $L$ states. The $H$ solution is featured by a defect line in the middle of the square, such that $\mathbf{n}$ is constant on either side of the defect line and has opposite orientations on both sides of the defect line. We label $H$ as a \emph{tessellating} solution, obtained as a juxtaposition of two $L$ solutions with opposite nematic directors. The line defect in the middle is avoidable in the tensor model, and thus this a \textit{"fake defect"} stemming from orientability issues. It remains to be seen whether such fake defects are of physical relevance or if they are a mere modelling artefact of vector-based models. The $C+$($-$) state has a central $+1$($-1$) point defect. In particular, the $C+$ state with a $+1$ point defect in the square centre is analogous to the Ring state  in the rLdG model, and is found in arbitrary 2D polygons \cite{han2020pol}. The ``$+$" refers to a circular profile, whereas the ``$-$" refers to a hyperbolic profile, and the signs in $N\pm$, $M\pm$, $X\pm$ mean the same thing.

At $\eta = 12$, the index of the $ISO$ solution increases to $6$ (Figure \ref{fig:2}(b)). We have new index-$5$ $LL$ and index-$4$ $X\pm$ states. The $LL$ state also has a central line defect as in the $H$ state, and can be viewed as a juxtaposition of two $L$ configurations on either side of the defect line, but $\mathbf{n}$ is orthogonal to the defect line in the $LL$ state whereas $\mathbf{n}$ is parallel to the defect line in the $H$ state.
The $X\pm$ state is composed of four different $D$ states on the four square quadrants, creating four point defects at the vertices accompanied by four point defects at the middle of the square edges and a $+1$($-1$) central point defect at the square centre. 

The index-$3$ $H$, index-$5$ $LL$, index-$4$ $X\pm$ states are all examples of tessellating solutions, suggesting that we can build new tessellating-type critical points by juxtaposing simpler building block-type critical points in sEL model, at least for our choice of the boundary conditions. As $\eta$ increases from $6$ to $12$, the index of $L$ states increases from $1$ to $2$. The index-$2$ $L$ connects to an index-$1$ $J$ solution, and the $J$ state acts as a transition state between the stable states $D$ and $R$. The $J$ state is similar to the $L$ state, except that the $J$ state is featured by a single line defect localised near a square edge, as opposed to a pair of line defects (low nematic order) near a pair of opposite square edges in the $L$ state.
More precisely, as $\eta$ increases $6$ to $12$, the index-$1$ $L$ state bifurcates into an index-$2$ $L$ state and an index-$1$ $R$ state, and the index-$1$ $R$ state further bifurcates into an index-$0$ $R$ state and an index-$1$ $J$ state. Similarly, as $\eta$ increases from $\eta=6$ to $\eta=12$ in the sEL model, the index of $C\pm$ decreases from $2$ to $0$, i.e., $C\pm$ becomes stable and some new critical points are observed --- the index-$1$ $M\pm$ and index-$2$ $N\pm$ saddle points. The $M\pm$ states are distinguished by localised regions of reduced order or point defects near the middle of a square edge, whereas the $N\pm$ states are distinguished by a localised region of low order near a square vertex. 


\begin{figure*}[htbp]
    \begin{center}
    \includegraphics[width=0.7\textwidth]{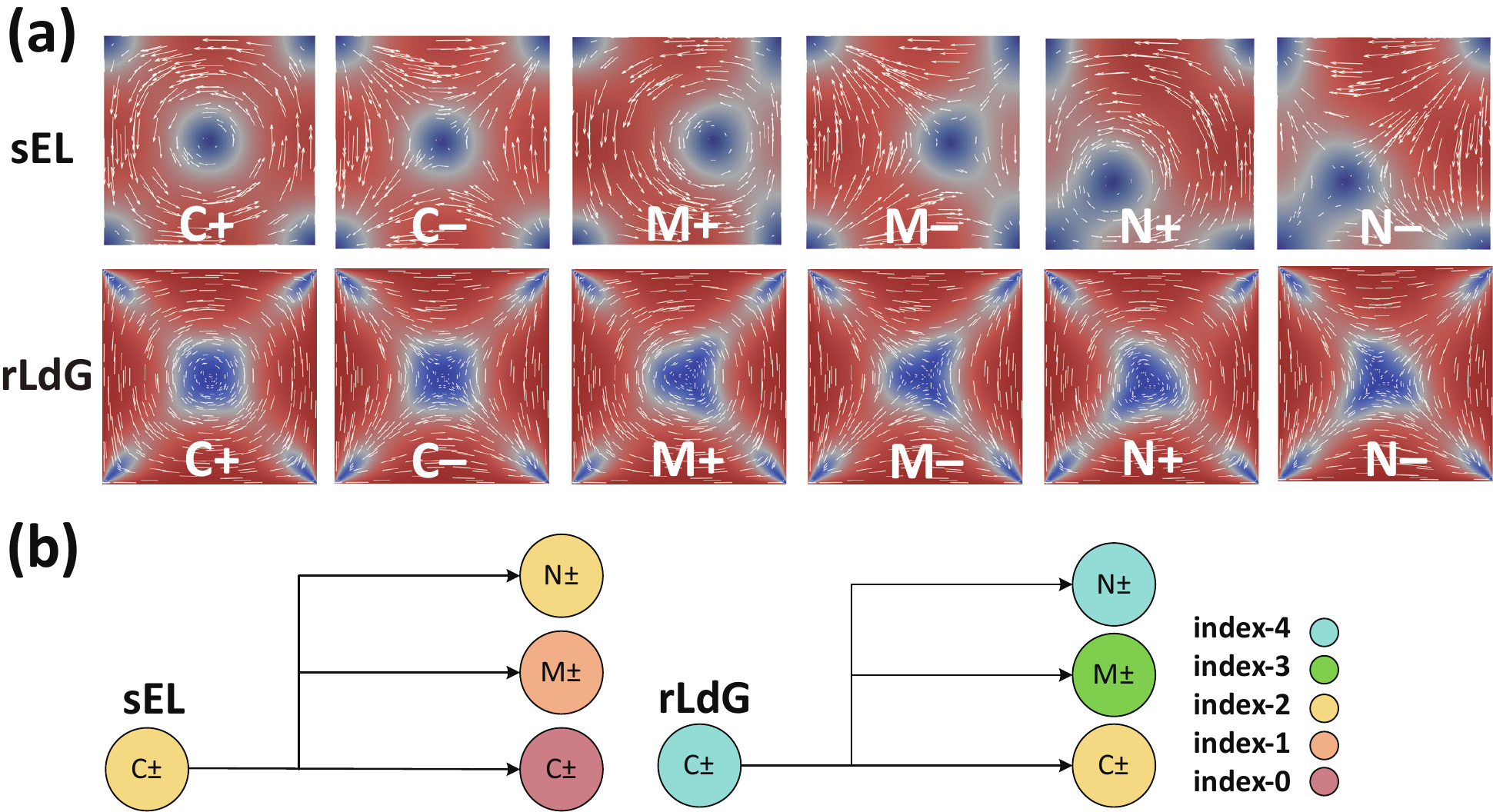}
    \caption{(a) The $C\pm$, $M\pm$, and $N\pm$ states in sEL model at $\eta = 12$ and in rLdG model at $\alpha = 50$. (b) The bifurcation from $C\pm$, to $C\pm$, $M\pm$, $N\pm$ states in sEL model as $\eta$ increases from $6$ to $12$ and in rLdG models as $\alpha$ increases from $45$ to $50$. The colors of the nodes specify the Morse indices of saddle points.}
    \label{fig:3}
    \end{center}
\end{figure*}
There are analogous $C\pm$, $M\pm$, and $N\pm$ states in the rLdG model too  \cite{yin2020construction}; see Figure \ref{fig:3}. The different interior defect profiles in the sEL and rLdG models can be attributed to the different boundary conditions, in particular, the unconstrained $s$ on $\partial \Omega$ in the sEL model. 

In rLdG model, as $\alpha$ increases from $45$ to $50$, the index-$4$ $C\pm$ saddle point bifurcates into an index-$4$ $N\pm$, an index-$3$ $M\pm$, and an index-$2$ $C\pm$ solutions, \cite{yin2020construction}. The differences between the $M\pm$ and $N\pm$ states are harder to spot in the rLdG model, except that the interior defect is more asymmetric and connected to a vertex in the $N\pm$ state, compared to the $M\pm$ state. The indices of $C\pm$, $M\pm$, and $N\pm$ in the rLdG model are higher than their counterparts in the sEL model, i.e., the $\pm 1$ point defect structure is more stable in the sEL model, as can be explained on the heuristic grounds.
The rLdG free energy prefers to split a $\pm 1$ point defect into two $\pm 1/2$ point defects, possibly along the two square diagonals, so that there are at least two directions of instability for the $C\pm$ state in the rLdG setting. For example, the planar radial state with a $+1$ central point defect is stable only if the domain size is extremely small, and the planar polar state with two $+1/2$-point defects along a circle diameter is the stable state for moderate or large disc sizes \cite{hu2016disclination}. However in the sEL model, interior half-integer point defects are disallowed and hence, these directions or modes of instability are absent, lending greater stability to interior vortices. We also note the absence of tessellating solutions in the rLdG setting. We would not expect to observe the $H$ or the $LL$ saddle points in the rLdG setting since these saddle points arise from the orientability of the sEL order parameter. However, the $X\pm$ saddle points may be observable in the rLdG setting with weaker boundary conditions e.g. with surface anchoring energies as opposed to Dirichlet tangent conditions in the rLdG framework.


\subsection{Large domain ($\eta = 20$)}\label{large}

\begin{figure*}[htbp]
    \begin{center}
    \includegraphics[width=\textwidth]{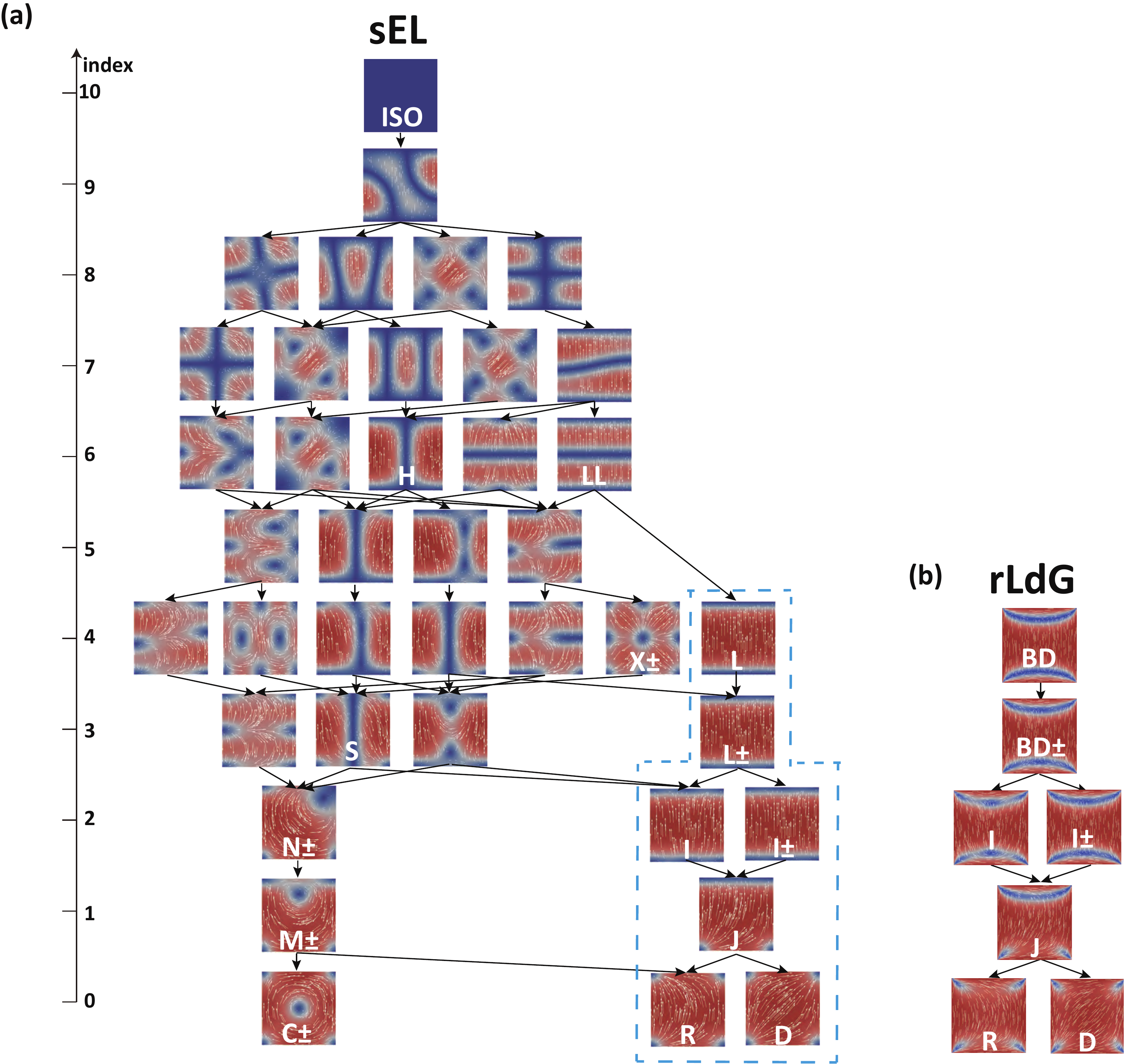}
    \caption{(a) The solution landscape of sEL model at $\eta = 20$. The height of a configuration corresponds to its Morse index. The sub-solution landscape boxed by blue dash has analogues in rLdG model shown in (b).}
    \label{fig:4}
    \end{center}
\end{figure*}

At $\eta = 20$, the sEL model (Figure \ref{fig:4}(a)) admits multiple new high-index solutions (from index-$3$ to index-$9$) with exotic configurations and complicated relationships between them. All the saddle points and minima for $\eta = 12$ in Figure \ref{fig:2} exist with possibly different indices. Most of new high-index solutions have regions of low order near the square edges that propagate into the interior, and separate differently ordered regions with different $\mathbf{n}$ profiles. These high-index saddle points arise from the relatively free boundary condition in the sEL model. Other high-index saddle points exhibit "fake defects" induced by the orientability of $\mathbf{n}$, and thus cannot be found in the rLdG model. Although the solution landscapes of the sEL model and rLdG model are quite different for large squares, there are some analogies. We find a complete sub-solution landscape (a part of solution landscape formed by critical points, with the index less than or equal to $4$) for which there are exact analogies between the sEL and rLdG models. In the dashed box in Figure \ref{fig:4}(a) and Figure \ref{fig:4}(b), the connections between the $L$, $L\pm$, $I$, $I\pm$,  , $J$, $D$ and $R$ states in the sEL model are analogous to the connections between the $BD$, $BD\pm$, $I$, $I\pm$, $J$, $D$ and $R$ states in the rLdG model. The index-$4$ $L$ or $BD$ connects to index-$3$ $L\pm$ or $BD\pm$, and further connects to index-$2$ $I\pm$ and $I$ solutions. The index-$2$ $I\pm$ and $I$ solutions connect the index-$1$ $J$ saddle point, which is the transition state between $D$ and $R$ stable states. In other words, the low-index saddle points in the sEL model are physically relevant in the sense that they survive in different models, and even with different boundary conditions.

The critical points: $L$, $L\pm$, $I$, $I\pm$, $J$, $D$ and $R$ in the sEL model (Figure \ref{fig:4}(a)), are analogous to the critical points: $BD$, $BD\pm$, $I$, $I\pm$, $J$, $D$ and $R$ in the rLdG model (Figure \ref{fig:4}(b)).
The $L\pm$ saddle point is composed of the top half of $L$ and bottom half of $I\pm$ i.e. there is an incipient half-integer point defect along one of the boundary defect lines/line of low order near a square edge. This can be seem more clearly in the analogous rLdG saddle point: $BD\pm$ which is composed of the top half of $BD$ and the bottom half of $I\pm$. In other words, there are two parallel line defects (of low order) near a pair of opposite square edges in $BD\pm$, and we see either a $+1/2$ or a $-1/2$-defect near the bottom line defect, and $\mathbf{n}$ is almost constant above this line defect. 
The $I$ or $I\pm$ saddle points in the rLdG model has $\pm 1/2$ interior defects near two opposite edges ($I$ has one $+1/2$ and one $-1/2$ defect, and $I\pm$ has a pair of $\pm 1/2$ defects), while the $I$ saddle point in the sEL model also has similar defective structures near pairs of opposite square edges. There are imprints of $\pm 1/2$-nematic defects in the sEL $I$ saddle point too, but the imprint is much stronger in the rLdG case owing to Dirichlet boundary conditions. In other words, we can view the sEL saddle points as relaxed versions of the rLdG saddle points. The $J$ saddle point has similar structure in both models.

\begin{figure*}[htbp]
    \begin{center}
    \includegraphics[width=0.7\textwidth]{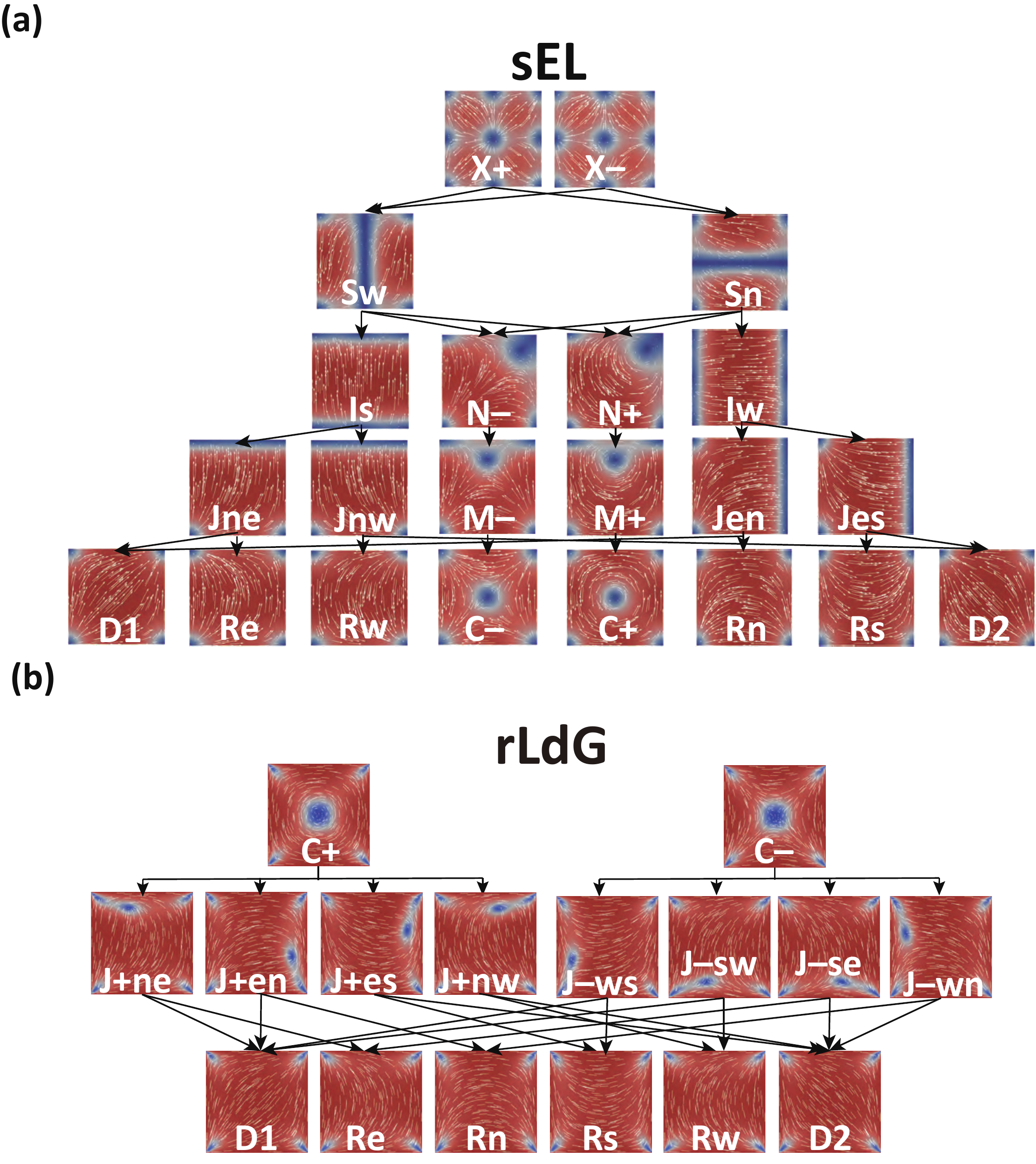}
    \caption{(a) Sub-solution landscape of sEL model starting from index-$4$ $X\pm$ at $\eta = 20$. (b) Sub-solution landscape of rLdG model starting from index-$2$ $C\pm$ at $\alpha = 75$.}
    \label{fig:6}
    \end{center}
\end{figure*}

In Figure \ref{fig:6}, we illustrate some dynamical pathways between saddle points in the sEL and rLdG models respectively. We look at dynamical pathways mediated by high-index saddle points, and transition pathways mediated by conventional index-$1$ saddle points. 
In the sEL model, all the stable states: $D1$, $Rw$, $Re$, $C-$, $C+$, $Rn$, $Rs$, and $D2$ can be connected by the index-$4$ $X\pm$ saddle point. The subscripts $n$, $s$, $w$, $e$ are short for ``north", ``south", ``west", and ``east", which correspond to up, down, left, and right in the readers' view. The sign ``$-$" or ``$+$" denotes either an $-1/2$ or an $+1/2$ interior point defect. For example, in $Rs$ the curved alignment points downwards; the $Jes$ saddle point has a low order line on the right and it connects with the $Rs$ state in the solution landscape; for the $Iw$ saddle point, the alignment is splay along the left edge; for the $J+es$ saddle point, the alignment is similar to the alignment of the $Jes$ state away from the low order edge on the right, accompanied by an interior $+1/2$ point defect.
In Figure \ref{fig:6}(a), one dynamical pathway from the index-$4$ $X+$ to the stable $Re$ state is $X+$ $\to$ $Sw$ $\to$ $Is$ $\to$ $Jne$ $\to$ $Re$; similarly, the transition pathway between $Re$ and $Rn$ can be mediated by mutiple index-$1$ transition states: $Re$ $\to$ $Jne$ $\to$ $D1$ $\to$ $Jen$ $\to$ $Rn$ or by high-index saddle points: $Re$ $\rightarrow$ $Jne$ $\rightarrow$ $Is$ $\rightarrow$ $Sw$ $\rightarrow$ $X+$ $\rightarrow$ $Sn$ $\rightarrow$ $Iw$ $\rightarrow$ $Jen$ $\rightarrow$ $Rn$. Within the limited scope of Figure \ref{fig:6}(a), we note that irrespective of the choice of the sub-parent state, $X+$ or $X-$, the dynamical pathway has access to both the index-$3$ saddle points, $Sw$ and $Sn$. Hence, one could have a similar dynamical pathway with different sub-parent states, at least for $X\pm$ in the sEL model.
For the rLdG model, all the stable states $D1$, $Rw$, $Re$, $Rn$, $Rs$, and $D2$ can be connected by index-$2$ $C\pm$. The index-$2$ $C+$ state connects with the stable states through the index-$1$ $J+$, with a $+1/2$ interior point defect. On the other hand, the dynamical pathways with the sub-parent state, $C-$, must proceed via $J-$, with a $-1/2$ interior point defect, to connect with the stable states. Hence, a reasonable conjecture is that dynamical pathways are more restricted to the high-index saddle points in the rLdG model, and there is less flexibility because of the strongly enforced Dirichlet boundary conditions.

\section{Discussions and conclusions}\label{sec:discussion}
We investigate the sEL model for NLCs on square domains, with tangent boundary conditions and low temperatures, and compare the sEL predictions with the rLdG predictions in \cite{yin2020construction}. We develop powerful solution landscape constructing algorithms that exploit the model symmetries, and these tools have applications beyond this paper. As a consequence, we can track both stable equilibria and unstable saddle points, and their connectivity as a function of square sizes in both models. The saddle points are classified by their Morse indices, and the algorithms can capture their indices and the unstable directions.

As mentioned in the Introduction, there are two notable differences between the sEL and rLdG models in our setting. Firstly, the sEL order parameter, $(s, \mathbf{n})$ is orientable whereas the rLdG order parameter is not orientable. Secondly, we use weak boundary conditions in the sEL setting which allows for a parent isotropic, $ISO$ solution or zero nematic order on the square edges. In contrast, we use Dirichlet boundary conditions with optimal nematic ordering on the square edges in the rLdG setting. For the 2D problem on a square, the two models reduce to two different boundary-value problems for the Ginzburg-Landau system of partial differential equations. In spite of the differences, both models have similarities for small and large squares. For small squares, the sEL model admits the unique $ISO$ state and the rLdG model admits the unique $WORS$ state. They serve as the parent states for all square sizes, and lose stability as the square size increases. In fact, we observe pitchfork bifurcations associated with the loss of stability of the $ISO$ ($WORS$) state in both cases. For sufficiently small squares, we observe analogous $L$ and $BD$ saddle points in the sEL and rLdG models respectively, followed by the stable $D$ states. As the square gets larger, the sEL solution landscapes get increasingly complicated. This is largely because of the orientable order parameter which introduces fake defects that separate regions with opposite $\mathbf{n}$-orientations, and because of novel tessellating solutions that arise from the weak boundary conditions e.g. the $H$, $LL$ and $X\pm$ saddle points. Further, without the non-orientability in the sEL model, we lose stable states with interior half-integer point defects and equally stabilise states with $\pm 1$-interior vortex structures ($C\pm$ with interior $\pm 1$ is a stable sEL state). The number of stable states is also doubled e.g. $4$ $D$ states and $8$ $R$ states. At $\eta=20$, we identify a sub-solution landscape for which there are exact correspondences between the sEL and rLdG models. Notably, in both cases, there are the stable $D$ and $R$ states for large squares, and the index-$1$ $J$ saddle point that acts as a transition state between the $D$ and $R$ states. However, the $C\pm$ saddle point has index $2$ in the rLdG setting, since there are at least two preferred directions of splitting the interior $\pm 1$-defect into a pair of $\pm 1/2$-defects along the square diagonals. This is a fascinating example of how stability properties of saddle points can be different for different choices of the model, and we speculate that the indices of saddle points with interior defects will always differ by $2$ on extremely large domain size, between the rLdG and sEL models respectively. We also make some preliminary remarks on transition pathways in the sEL framework, for large squares, which seem to suggest that the transition pathways are more flexible or less sensitive to the choice of the parent state in the sEL problem. We conjecture that the greater flexibility can be attributed to the weak boundary conditions, which naturally enhance the number of connections between the saddle points and the stable states in the sEL problem.

There are numerous open questions and our work is an informative forward step for comparing different models for NLCs in confinement. We see that while the sEL and rLdG models have similar stable states, the connecting saddle points can be very different. Hence, from a physical standpoint, these models can predict very different switching mechanisms mediated by different types of saddle points. A very natural extension of this work would be to compare this sEL problem with the rLdG problem, such that the rLdG problem has weakly enforced tangent boundary conditions, i.e. if we allow for relaxed order on the square edges in the rLdG setting. In this case, we certainly expect to see stronger similarities between the saddle points, for larger squares, in the sEL and rLdG settings. However, the boundary-value problems will still  be different, even with weak boundary conditions, for both the sEL and rLdG problems. Therefore, the solution landscapes will not be identical in this case, but stronger overlaps can be expected since boundary defects will be allowed in the rLdG setting too. This will allow us to study the challenging interplay between the choice of the order parameter, the model and the boundary conditions in the solution landscapes. 

A further observation pertains to how the mathematical model stabilises or de-stabilises certain defects. Recent work in the rLdG setting shows that elastic anisotropy can stabilise $+1$-interior point defects on square domains \cite{han+majumdar+harris+zhang+2021}. Further, models with nemato-magnetic coupling can also stabilise interior nematic vortices on square domains in the rLdG setting \cite{hanpre2021}. In this paper, the choice of the sEL order parameter and the boundary conditions stabilises interior vortices, so this raises the hugely challenging question of whether we can identify and quantify the different effects that stabilise nematic defects. If successfully done, this could be profoundly important for both fundamental science and applications.
Equally importantly, this approach can also be applied to general mathematical models beyond liquid crystal theory. 
For instance, the Ohta--Kawasaki model \cite{Takao1986Equilibrium} and the Landau--Brazovskii model \cite{S1975Phase} are phase field models that can describe the diblock copolymer, derived from particle-based models by mean-field approximations \cite{Fredrickson2006polymer, 2007Field}. The comparison between these two models, either theoretically or numerically, is interesting but challenging. We believe that our solution landscape approach that delves into the connectivity of unstable and stable states could be valuable in this context too. We hope to pursue some of these challenging research avenues in future work.
\section*{Acknowledgments}
L. Zhang was supported by the National Natural Science Foundation of China No. 12050002. L. Zhang and A. Majumdar acknowledge the support from Royal Society Newton Advanced Fellowship. A.Majumdar acknowledges support from the University of Strathclyde New Professor's Fund and a Leverhulme International Academic Fellowship. A.Majumdar also acknowledges support from an OCIAM Visiting Fellowship at the University of Oxford, a visiting professorship at the University of Bath and IIT Bombay. Y. Han acknowledges the support from Royal Society Newton International Fellowship. J. Yin acknowledges the support from the Elite Program of Computational and Applied Mathematics for Ph.D. Candidates of Peking University.

\bibliographystyle{unsrt}

\end{document}